\documentclass[10pt,letterpaper,twocolumn]{article}
\usepackage{ol}
\usepackage{amsmath}
\usepackage{bm}
\begin{document}

\twocolumn[ 

\title{Goos-H{\"a}nchen induced vector eigenmodes in a dome cavity}

\author{David H. Foster} 

\address{Deep Photonics Corp., 5121 SW Hout St., Corvallis, OR, 97333}

\author{Andrew K. Cook and Jens U. N\"{o}ckel}

\address{Department of Physics, University of Oregon, 1371 E 13th Avenue,
Eugene, OR, 97403}

\begin{abstract}
We demonstrate numerically calculated electromagnetic eigenmodes of a 3D dome cavity resonator that owe their shape and character entirely to the Goos-H\"{a}nchen effect. 
The V-shaped modes, which have purely TE or TM polarization, are well described by a 2D billiard map with the Goos-H\"{a}nchen shift included. 
A phase space plot of this augmented billiard map reveals a saddle-node bifurcation; the stable periodic orbit that is created in the bifurcation corresponds to the numerically calculated eigenmode, dictating the angle of its ``V''. 
A transition from a fundamental Gaussian to a TM V mode has been observed as the cavity is lengthened to become nearly hemispherical.
\end{abstract}


] 


\noindent The Goos-H\"{a}nchen (GH) effect\cite{GoosAPL1947}, which typically occurs upon reflection from a dielectric interface or layered structure, is an apparent parallel translation of the reflected beam away from its geometrically expected position.
A simple treatment\cite{ArtmannAPL1948} shows that the size of the GH shift is proportional to the derivative of the reflection phase with respect to the angle of incidence.
In addition to shifting beam position, the the GH effect can manifest itself in alterations of differential cross sections\cite{TranOL1995}, of laser mode dynamics\cite{DutriauxJOSAB1992}, and of mode spectra~\cite{ChowdhuryJOSAA1994}. 

Here we present and analyze a class of eigenmodes in a non-paraxial dome cavity, the ``V modes'', showing that the modes are straightforwardly explained by a stable periodic orbit in a billiard map that is augmented by the Goos-H\"{a}nchen effect. A ray analysis of the cavity geometry {\em without} taking the Goos-H\"{a}nchen shift into account fails to predict these modes. 
To the authors' knowledge, this Letter presents the first demonstation of the GH effect, or any nonspecular effect\cite{NasalskiJOSAA1996}, creating entirely new eigenmode patterns.

Our model dome cavity consists of a perfectly conducting concave spherical mirror, M1, facing a planar thin film (distributed Bragg reflector), denoted by M2.
The electromagnetic eigenmodes are calculated by the basis expansion methods which are fully described in Ref.~\onlinecite{FosterOC2004methods}. 
We first described the V modes in Ref.~\onlinecite{FosterOC2004methods}, but no physical explanation for their appearance has been put forward until now. 

\begin{figure}[h]
\centerline{\includegraphics[width=\columnwidth]{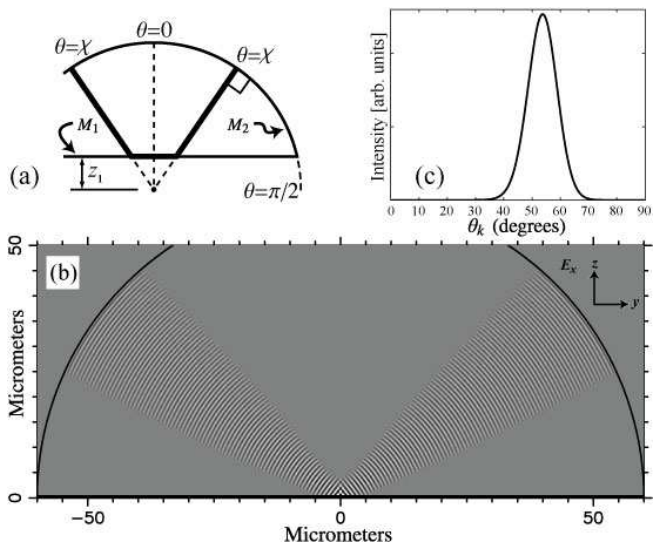}}
\caption{\label{fig: Vmode pic}(a) Diagram of the simple ray model for the V mode, in a dome cavity, defining the polar angle $\theta$ and the mirror labels $M_1$, $M_2$. (b) Gray-scale plot of a VTE$_0^1$ mode at mirror offset $z_1 =0.4\,\mu$m. Parts of the dome surface are outside the figure for space reasons. (c) Angular spectrum of this mode.}
\end{figure} 
    To provide quantitative results, we fix some parameters at the outset, cf. Fig.~\ref{fig: Vmode pic}(a): 
The radius of M1 is $R=60\,\mu$m and the length of the resonator, measured from the front surface of M2 to the zenith of M1, is $L=R-z_1$, where we will focus on the range $0.3 \leq z_1 \leq 0.7\,\mu$m. This means the resonator is nearly hemispherical, and increasing mirror offset $z_1$ makes the cavity more paraxial. 
The layer structure of M2 is $IAAAA(BA)^{22}O$ where $I$ denotes the vacuum region inside the cavity, each $A$ denotes a layer of $n=3.52$ material (AlGaAs) of quarter wave thickness at $570\,$nm, each $B$ denotes a layer of $n=3.0$ material (AlAs), also of quarter wave thickness at $570\,$nm, and $O$ denotes the vacuum region outside the cavity.

Axial symmetry allows one to label all cavity modes by a total angular momentum quantum number $m$; 
the complex electric and magnetic fields $E_{\rho}$, $E_{\phi}$, $E_z$, $B_{\rho}$, $B_{\phi}$, and $B_z$ (in cylinder coordinates) are proportional to $\exp (i m \phi)$. In this paper, we present data only for modes with $m=\pm1$.
The vectorial eigenmodes typically found in an axially symmetric cavity are in general neither TE nor TM. Here, TE refers to fields whose angular spectrum contains only s-polarized plane waves with respect to the Bragg mirror; TM denotes a p-polarized plane wave decomposition. 

Fig.~\ref{fig: Vmode pic}(b) shows a V mode having wavelength $\lambda=755.44\,$nm at $z_1=0.4\,\mu$m.
The contrast-enhanced plot shows the $y$-$z$ slice of the physical $E_x$ field ($z$ is the dome axis). 
Azimuthally, the electric field here is proportional to $\exp[i (\pm \phi - \omega t)]$, which may be combined to make the ``linearly'' polarized versions $\cos\!{\rm /}\!\sin(\phi) \exp(-i \omega t)$. The V mode is almost completely TE polarized. 
Fig.~\ref{fig: Vmode pic}(c) shows the s-polarized (TE) polar angle distribution of plane waves, $|\psi_s(\theta_k)|^2$, for this mode; $\theta_k$ is the angle of incidence of a plane wave with respect to the cavity axis. The p-polarized admixtures are negligible on the scale of the plot.

Averaging $\theta_k$ against this distribution yields $\langle \theta_k \rangle = 53.6^{\circ}$, which one can identify as the ``opening angle'' of the V-shaped beam. 
Similar modes have been observed which are entirely TM. We also found TE modes which exhibit a transverse node in each side of the V. This leads us to introduce the notation VTM$_{\nu}^m$ and VTE$_{\nu}^m$ for these V modes, where $m$ is the total angular momentum and $\nu$ is  the number of nodes in the transverse wavefunction of each side of the V.
The mode in Fig.~\ref{fig: Vmode pic}  is then VTE$_0^1$. 

We now turn to the explanation of these V modes in terms of the Goos-H\"{a}nchen effect: 
The GH shift \emph{along the reflecting surface}, of an s- or p-polarized beam is\cite{ArtmannAPL1948}
\begin{equation}
\Delta_{s/p} (\theta_k; k) = - \frac{1}{k \cos \theta_k} \frac{{\rm d}\phi_{s/p} (\theta_k; k)}{{\rm d}\theta_k}, \label{eqn: parallel GH shift}
\end{equation} 
where $k=2 \pi / \lambda$ and $\phi_{s/p}$ is the plane wave reflection phase of the surface. 
A simple self-retracing periodic orbit which reflects perpendicularly from M2  
 can exist at $\theta_k = \chi > 0$ if the GH shift equals the geometrically determined segment, that is if
\begin{equation}
\Delta_{s/p} (\chi; k)= 2 z_1 \tan \chi , \label{eqn: PO condition}
\end{equation}
where the right hand side is seen to be the length of the horizontal segment of the orbit shown (bold line) in Fig.~\ref{fig: Vmode pic}(a). 
 
\begin{figure}[h]
\centerline{\includegraphics[width=\columnwidth]{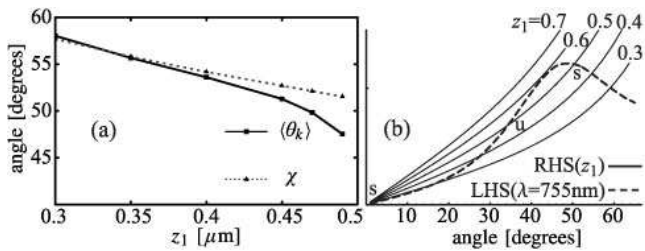}}
\caption{\label{fig: 2}(a) Comparison of $\langle \theta_k \rangle$ with $\chi$ as a function of $z_1$ for a VTE$_0^1$ mode. 
The $\chi_s$ calculation takes its $k$ values from the actual modes.
(b) LHS, RHS of Eq.~(\ref{eqn: PO condition}) (using $\Delta_s$) with respect to $\chi$ (deg), with fixed $\lambda$. 
The solutions corresponding to (s)table and (u)nstable orbits alternate.}
\end{figure}
Since $z_1$ is given physically and $k$ may be coarsely chosen, Eq.~(\ref{eqn: PO condition}) is in practice solved for $\chi_s$ or $\chi_p$ to predict the angle of the V modes.
The prediction for the VTE$_0^1$ mode shown in Fig.~\ref{fig: Vmode pic} is $\chi_s = 54.2^{\circ}$, in excellent agreement with the average angle $\langle \theta_k \rangle = 53.6^{\circ}$ obtained for the numerically calculated eigenmode. 
Numerical experimentation shows that $\chi_{s/p}$ increases to near $90^{\circ}$ as $z_1 \rightarrow 0_{+}$.
Fig.~\ref{fig: 2}(a) shows good agreement between $\chi_s$ and $\langle \theta_k \rangle$ for a VTE$_0^1$ mode followed from $z_1=0.3$ to $z_1=0.49$. 
(Following the VTE$_0^1$ mode pattern over this range required ``hopping'' across mode anticrossings at $z_1=0.46$ and $z_1=0.48$.) 

\begin{figure}[h]
\centerline{\includegraphics[width=0.75\columnwidth]{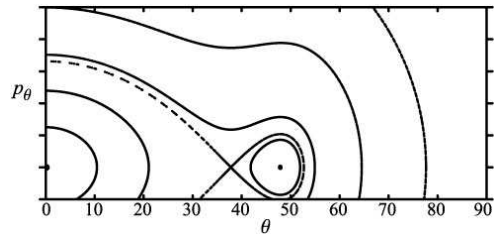}}
\caption{\label{fig: 3}Billiard map with GH shift for TE polarization at $z_1=0.55$, showing creation of the stable V-ray island and separatrix.}
\end{figure}
To characterize the conditions under which the V modes are stable, it is useful to investigate the phase-space of the classical ray dynamics. The ray orbit shown in Fig.~\ref{fig: Vmode pic}(a) is a {\em periodic orbit} (PO) of a billiard map which obeys specular reflection at the top mirror M2, but at M1 receives a GH shift, along the M1 surface, the length of which is given by Eq.~(\ref{eqn: parallel GH shift}). 
To specify this map, we follow a ray through multiple reflections, and just before the $n$-th bounce at mirror M2, note its  polar coordinate $\theta^{(n)}$ as shown in Fig.~\ref{fig: Vmode pic}(a), as well as the component $p_{\theta}^{(n)}$ of the (unit) momentum in the $\hat{\bm\theta}$ direction. The billiard map $T$ maps this pair of ray properties from one reflection at M2 to the next one: 
\begin{equation}
T : (p_{\theta}^{(n)}, \theta^{(n)}) \mapsto (p_{\theta}^{(n+1)}, \theta^{(n+1)}),  \label{eq:billiardmap}
\end{equation}
A plot of nine trajectories, each of 1000 bounces, of the map $T$ for $z_1=0.55$ and s-polarization (TE) is shown in Fig.~\ref{fig: 3}.
Such a phase space plot is very much like the Poincar\'{e} section of a dynamical system and may be analyzed as such. 
The island structure surrounding the stable $\theta=0$ PO manifests itself in wave physics as the family of paraxial Gaussian modes.
The separatrix indicates an unstable PO at $\theta \approx 37^{\circ}$ and the second island indicates a stable PO at $\theta \approx 49^{\circ}$. 
As shown in Fig.~\ref{fig: 2}(b), these two POs are created at the same point (this is called a saddle-node bifurcation) as $z_1$ is lowered to about $0.6$, and move apart in $\theta$ as $z_1$ is further decreased.
As $z_1$ decreases, the $\theta=0$ island shrinks, while the new island grows.
As it grows, one or more existing wave modes may become distorted until one of them becomes the V mode; the V modes live inside the new island. 
If the GH effect is turned off in the billiard map (\ref{eq:billiardmap}), there is no new island, and the only PO that exists is the $\theta=0$ orbit. This leads to the important conclusion that {\em only a ray map augmented by the GH effect} predicts the stable cavity modes correctly. This analysis can be applied equally to the s- or p-polarized V modes, by inserting the appropriate reflection phase  $\phi_{s/p}$ in the map $T$.
 				 
At sufficiently large mirror offset  $z_1$, we should be able to make contact with Gaussian-beam theory.
The resonator modes may be characterized by their transverse field patterns, which are described in Refs.~\onlinecite{YuIEEETMTT1984, BliokhOC2005, FosterOC2004methods, FosterOL2004, FosterThesis}. 
In Refs.~\onlinecite{FosterOC2004methods, FosterOL2004, FosterThesis}, a dielectric mirror provides a form birefringence which causes the transverse mode patterns of many typical eigenmodes to be symmetric and antisymmetric mixtures of specific pairs of vectorial Laguerre-Gauss modes, causing either TE or TM polarization to dominate the plane-wave spectrum. 
Starting from paraxial theory, these ``mixed'' modes can be denoted by the transverse order $N$, the angular momentum $m$, and whether or not they are quasi-TE or quasi-TM:
\begin{equation}
{\rm QTE(M)}_N^m = {\rm LG}_{(N-|m-1|)/2}^{m-1} \bm{\sigma}^{+}
\pm {\rm LG}_{(N-|m+1|)/2}^{m+1} \bm{\sigma}^{-}
\end{equation}
for $|m| \geq 1$, $N-|m+1| \in \{0, 1, \ldots\}$, and $N-|m-1| \in \{0, 1, \ldots\}$.
Here $\bm{\sigma}^{\pm}$ is the Jones vector for right/left circular polarization and LG denotes the Laguerre-Gauss wave pattern\cite{FosterOL2004}. A detailed discussion of the mixed modes is given in Ref.~\onlinecite{FosterThesis};
they, along with pure Gaussians occuring for $\vert m\vert=N+1$, comprise the background of 
eigenmodes against which the V modes appear.

\begin{figure}[h]
\centerline{\includegraphics[width=\columnwidth]{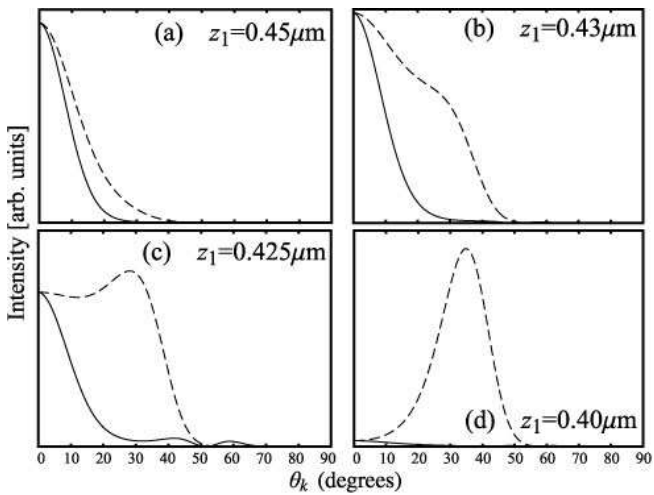}}
\caption{\label{fig: VTM creation}Angular spectra showing the creation of the VTM$_0^1$ mode from the fundamental Gaussian as $z_1$ decreases. 
Dashed: $|\psi_s|^2$, solid: $|\psi_p|^2$.}
\end{figure}
We numerically observe that V modes evolve from some existing mode of this paraxial type as $z_1$ is reduced and the stable island grows. As an example of this metamorphosis, a VTE$_0^1$ mode was created from the predominantly TE mixed mode having the tranverse E field, in the paraxial limit, given by ${\rm QTE}_4^1 \equiv {\rm LG}_2^0 \bm{\sigma}^{+}$$+{\rm LG}_1^2 \bm{\sigma}^{-}$. On the other hand, as shown in Fig.~\ref{fig: VTM creation}, a VTM$_0^1$ mode was created from the fundamental Gaussian mode TEM$_{00}$, which originally has circular polarization. The polarization-dependent stabilization provided by the GH effect thus overcomes the polarization state expected for the paraxial fundamental mode in an axially symmetry cavity. This non-paraxial nature of the V modes underlines the predictive power of the Goos-H{\"a}nchen picture.

Although the numerical results presented here were limited to $|m|=1$, we have observed VTM$_0^2$, VTE$_0^2$, and VTE$_1^2$, and expect V modes to generally exist for $|m| \geq 1$. Other dielectric stack designs than the one described here have also been observed to induce V modes. 

We note that patterns reminiscent of the V modes appear in an earlier work by Bl{\"u}mel {\em et al.}\cite{BlumelPRL1996} which discusses scalar waves in a purely two-dimensional circle with a dielectric boundary across its diameter. This roughly corresponds to the limit $z_1\to 0$ in our model, and neither the question of the GH effect nor the $z_1$ dependent opening angle of the V modes are addressed in that work. As pointed out in Ref.~\onlinecite{BlumelPRL1996}, semiclassical techniques that may be appropriate for smoothly varying media or simple boundary conditions (e.g., Dirichlet) need to be extended in the presence of dielectric interfaces. The results presented here were obtained in a simpler and physically appealing way, by adding nonspecular corrections to the classical ray mapping $T$. 

The V modes demonstrate that small corrections to specular reflection, such as the Goos-H\"{a}nchen shift, can have large qualitative effects.
Billiard models augmented by nonspecular corrections possess new phase space structures which correspond to the new modes. 
The V mode in particular provides a potential lasing mode with two or four entrance/exit channels, yet it economically uses only one curved surface. Interestingly, the spreading angle of each limb of the V modes is {\em smaller} than predicted by a Gaussian-beam calculation based on the path length of the V including GH shift. In future work, we will investigate to what extent a stability analysis of the ray map $T$ is able to explain this phenomenon.

This work was supported by NSF Grant ECS-0239332. 



\end{document}